# Approximate Analytical Solutions to the Relativistic Isothermal Gas Spheres

A. S. Saad[1,2], M. I. Nouh[1,3], A. A. Shaker[1] and T. M. Kamel[1]

[1]Department of Astronomy, National Research Institute of Astronomy and Geophysics, Cairo, Egypt
Email: saad6511@gmail.com
[2]Department of Mathematics, Preparatory Year, Qassim University, Buraidah, KSA
[3]Department of Physics, College of Science, Northern Border University, Arar, KSA
Email: abdo_nouh@hotmail.com

**Abstract**

In this paper, we introduce a novel analytical solution to Tolman-Oppenheimer-Volkoff (TOV) equation, which is ultimately a hydrostatic equilibrium equation derived from the general relativity in the framework of relativistic isothermal spheres. Application of the traditional power series expansions on solving TOV equation results in a limited physical range to the convergent power series solution. To improve the convergence radii of the obtained series solutions, a combination of the two techniques of Euler-Abel transformation and Padé approximation has done. The solutions are given in $\xi$-$\theta$ and $\xi$-$v$ phase planes taking into account the general relativistic effects $\sigma$ = 0.1, 0.2 and 0.3. An Application to a neutron star has done. A Comparison between the results obtained by the suggested approach in the present paper and the numerical one indicates a good agreement with a maximum relative error of order $10^{-3}$, which establishes the validity and accuracy of the method. The procedure we have applied accelerated the power series solution with about ten times than of traditional one.

## 1. Introduction

Isothermal models play an important role in many astrophysical problems, in particular in stellar structure and galactic dynamics (Chandrasekhar 1942; Binney &Tremaine 1987; Rose 1998; Chavanis 2002, 2008). As for stellar structure and evolution theory, one can obtain the behavior of the physical variables through the isothermal self-gravitating spheres. In the equation of state $P = K\rho^{1+1/n}$ (where $P$ is the pressure, $\rho$ is the density and $K$ is the polytropic constant), the polytropic index $n$ ranges from 0 to $\infty$, Liu (1996) and Mirza (2009). In galactic dynamics, the polytropic index $n$ is larger than 0.5 and the Lane-Emden equation for isothermal configurations is considered as a generating function of potential models for flattened galactic systems (Binney & Tremaine 1987).

In the framework of the Newtonian mechanics (non-relativistic), isothermal spheres have been much investigated by many authors (Milgrom 1984, Liu 1996; Natarajan and Lynden-Bell, 1997; Roxburgh and Stockman, 1999; Hunter 2001; Nouh, 2004; Mirza, 2009; Raga et al. 2013a, b). In the framework of the theory of general relativity, some numerical studies on the relativistic isothermal spheres have performed by several authors (Edward & Merilan 1981a, b; Zhang e al. 1983; Chavanis 2002 and 2008). Chau et al. (1984) extended the work done by Edward & Merilan (1981a, b) to determine the static structure of general relativistic, partially degenerate, isothermal (with arbitrary temperature) configurations. Chavanis (2002, 2008 and 2009) investigated the structure and stability of isothermal gas spheres in the framework of general relativity with a linear equation of state $P = q \in$ ($\in$: the mass-energy density, $P$: pressure and $q \rightarrow 0$ for Newtonian gravity). Sharma (1990) introduced approximate analytical solutions to TOV equation of hydrostatic equilibrium in the concise and simple form using Pade approximation technique. His results show that general relativity isothermal configurations are of finite extent. The geometrical size derived by Sharma (1990) $\xi = \xi_1$ of the configurations is limited to $\xi_1 = 2$.

Due to the lack of a full analytical solution to TOV equation of hydrostatic equilibrium and the importance of the study of relativistic isothermal configurations to various astronomical objects, we introduce a new approximate analytical solution to Tolman-Oppenheimer-Volkoff (TOV) equation based on a combination of the two techniques of Euler-Abel transformation and Padé approximation (Nouh & Saad 2013). The constructed analytical solution is in a general form such that we can increase the number of terms of the series solution for the desired accuracy. We display the solution curves in $\xi$-$\theta$ and $\xi$-$\upsilon$ phase planes for three general relativistic effects $\sigma = 0.1$, 0.2 and 0.3. The proposed method is promising and can examine efficiently the behavior of the physical parameters of stars, such as density, pressure, temperature, etc. from the center towards the surface.

## 2. Formulations

The Tolman-Oppenheimer-Volkoff equation of the isothermal gas sphere could be given by (Sharma, 1990)

$$\frac{dP}{dr} = -\left(\frac{GM_r}{r^2}\right)\frac{\rho + (P/c^2)(1 + (4\pi P r^3 / M_r c^2))}{(1 - (2GM_r / rc^2))}, \tag{1}$$

where $M_r$ is the total mass energy or 'effective mass' of the star of radius $r$ including its gravitational field - i.e,

$$M_r = \frac{4\pi}{c^2}\int_0^r \rho c^2 r^2 dr. \tag{2}$$

Define the variables, $\xi$, $\theta$, $\upsilon$ and the relativistic parameter $\sigma$, respectively by the following Equations.

$$\xi = rA;\ \rho = \rho_c e^{-\theta};\ M_r = \frac{4\pi\rho_c}{A^3}\upsilon(\xi);\ A^2 = 4\pi G\rho_c/\sigma c^2,\tag{3}$$

$$\sigma = \frac{P_c}{\rho c^2},\tag{4}$$

where $\rho_c$ and $P_c$ define the central density and central pressure of the star respectively. $\upsilon(\xi)$ is the mass function of radius $\xi$, and $c$ is the speed of light. Equations (1) and (2) can be transformed into the dimensionless forms in $(\xi,\theta)$ plane as

$$\frac{(1-2\sigma\upsilon(\xi)/\xi)}{1+\sigma} + \xi^2\frac{d\theta}{d\xi} - \upsilon(\xi) - \sigma e^{-\theta}\xi^3 = 0,\tag{5}$$

$$\frac{d\upsilon}{d\xi} = \xi^2 e^{-\theta},\tag{6}$$

which satisfy the initial conditions

$$\theta(0) = 0;\ \frac{d\theta(0)}{d\xi} = 0;\ \upsilon(0) = 0.\tag{7}$$

If the pressure is much smaller than the energy density at the center of a star (i.e. $\sigma$ tends to zero), the non-relativistic case, then the systems (5) and (6) reduce to yield the classical nondegenerate isothermal structure equations (cf. Chandrasekhar 1942)

$$\frac{1}{\xi^2}\frac{d}{d\xi}(\zeta^2\frac{d\theta}{d\xi}) = e^{-\theta}.\tag{8}$$

### 3. Series Solution

Equation (5) can be written in the form

$$\xi^2\frac{d\theta}{d\xi} - 2\sigma\upsilon\xi\frac{d\theta}{d\xi} - \upsilon - \sigma\upsilon - \sigma e^{-\theta}\xi^3 - \sigma^2 e^{-\theta}\xi^3 = 0.\tag{9}$$

Consider a series expansion near the origin $\xi = 0$, of the form

$$\theta(\xi) = \sum_{k=1}^{\infty} a_k \xi^{2k},\tag{10}$$

that satisfies the initial conditions (7); at $\theta(0) = 0$ yields $a_0 = 0$,

then

$$\frac{d\theta}{d\xi} = \sum_{k=1}^{\infty} 2k a_k \xi^{2k-1}, \tag{11}$$

$$\xi \frac{d\theta}{d\xi} = \sum_{k=1}^{\infty} 2k a_k \xi^{2k}, \tag{12}$$

Multiply Equation (11) by $\xi^2$ yields

$$\xi^2 \frac{d\theta}{d\zeta} = \sum_{k=1}^{\infty} 2k a_k \xi^{2k+1}. \tag{13}$$

With the help of algebraic operations on series (Nouh and Saad 2013), we get

$$e^{-\theta} = \sum_{k=0}^{\infty} \alpha_k \xi^{2k}; \quad \alpha_0 = e^{(-a_0)}; \quad \alpha_k = -\frac{1}{k} \sum_{i=1}^{k} i a_i \alpha_{k-i}, \quad \forall k \geq 1. \tag{14}$$

Inserting Equation (9) in Equation (6) yields

$$\frac{d\upsilon}{d\xi} = \xi^2 \sum_{k=0}^{\infty} \alpha_k \xi^{2k} = \sum_{k=0}^{\infty} \alpha_k \xi^{2k+2}. \tag{15}$$

Integrating both sides of the last equation gives

$$\upsilon = \int \sum_{k=0}^{\infty} \alpha_k \xi^{2k+2} d\xi = \sum_{k=0}^{\infty} \frac{\alpha_k}{2k+3} \xi^{2k+3}. \tag{16}$$

Multiply both sides of sides of Equation (12) by $\upsilon$, then using Equation (16) obtain

$$\xi \frac{d\theta}{d\zeta} \upsilon = \left( \sum_{k=0}^{\infty} 2(k+1) a_{k+1} \xi^{2k+2} \right) \left( \sum_{k=0}^{\infty} \frac{\alpha_k}{2k+3} \xi^{2k+3} \right). \tag{17}$$

Let

$$f_k = 2(k+1) a_{k+1}, \text{ and } g_k = \frac{\alpha_k}{2k+3}, \tag{18}$$

then

$$\xi \frac{d\theta}{d\zeta} \upsilon = \left( \sum_{k=0}^{\infty} f_k \xi^{2k+2} \right) \left( \sum_{k=0}^{\infty} g_k \xi^{2k+3} \right). \tag{19}$$

Implementing the formula of multiplication of two series (Nouh & Saad 2013), Equation (19) becomes

$$\xi \frac{d\theta}{d\zeta} \upsilon = \left( \sum_{k=0}^{\infty} f_k \xi^{2k+2} \right) \left( \sum_{k=0}^{\infty} g_k \xi^{2k+3} \right) = \sum_{k=0}^{\infty} \gamma_k \xi^{2k+5}, \tag{20}$$

where

$$\gamma_k = \sum_{i=0}^{k} f_i g_{k-i}. \tag{21}$$

Substituting Equations (13), (14), (16) and (20) in Equation (9) yields

$$\sum_{k=1}^{\infty} \left[ 2(k+1)a_{k+1} - 2\sigma\gamma_{k-1} - \frac{\alpha_k}{2k+3} - \sigma \frac{\alpha_k}{2k+3} - \sigma\alpha_k - \sigma^2 \alpha_k \right] \xi^{2k+3} = 0, \tag{22}$$

that is

$$2(k+1)a_{k+1} - 2\sigma\gamma_{k-1} - \frac{\alpha_k}{2k+3} - \sigma \frac{\alpha_k}{2k+3} - \sigma\alpha_k - \sigma^2 \alpha_k = 0.$$

Then, the series coefficients formulation has the form

$$a_{k+1} = \frac{1}{2(k+1)} \left[ 2\sigma\gamma_{k-1} + \frac{\alpha_k}{2k+3} + \sigma \frac{\alpha_k}{2k+3} + \sigma\alpha_k + \sigma^2 \alpha_k \right], \tag{23}$$

where

$$\gamma_{k-1} = \sum_{i=0}^{k-1} f_i g_{k-i-1}; \quad \alpha_k = -\frac{1}{k} \sum_{i=1}^{k} i a_i \alpha_{k-i}; \quad \forall k \geq 1; \quad \alpha_0 = e^{(-a_0)}; \quad a_0 = 0. \tag{24}$$

As for example, the coefficient $a_1$ can be computed as

$$a_1 = \frac{1}{2} \left[ \frac{1}{3} + \frac{\sigma}{3} + \sigma + \sigma^2 \right] = \frac{1}{6}(1+\sigma)(1+3\sigma). \tag{25}$$

Consider $\sigma = 0$ in Equation (18), we get $a_1 = \frac{1}{6}$ which is the first series term in the case of the non-relativistic Lane-Emden solution.

## 4. Results

The analytical solution of Equations (5) and (6) with the given initial conditions in Equation (7) determines the relativistic structure of the configuration. This solution is represented by Equation (10) together with Equations (23) to (25). The radius of convergence $\xi_1$ of the power series solution without applying any acceleration techniques is limited. Tables 1, 2 and 3 show the solutions $\theta(\xi_1)$, the mass function $v(\xi_1)$ and their errors between analytical and numerical solutions $\varepsilon_1$ and $\varepsilon_2$ respectively. It is found that the maximum radii of convergence for the relativistic effects $\sigma = 0.1$, $0.2$ and $0.3$ are 3.58, 3.23 and 2.8 respectively. Beyond the mentioned values, the power series solution is either slowly convergent or divergent. It is worth noting that we have used the fourth order Runge-Kutta method for numerical solution of TOV equation.

Table 1. Radius of convergence of the power series solution of TOV equation for the relativistic effect $\sigma = 0.1$ before doing acceleration.

| $\xi$ | $\theta(\xi_1)$-An | error ($\varepsilon_1$) | $v(\xi_1)$-An | error ($\varepsilon_2$) |
|---|---|---|---|---|
| 0.1 | 0.0023821620 | -1.0260393175E-006 | 0.0003328572 | 3.7246274503E-009 |
| 0.3 | 0.0213554534 | -8.7300804448E-007 | 0.0088854114 | 2.7700080786E-008 |
| 0.5 | 0.0588588473 | -8.2433308801E-007 | 0.0402206825 | 4.3356597944E-008 |
| 0.7 | 0.1140289083 | -7.7062136704E-007 | 0.1067702316 | 6.2923817407E-008 |
| 1.0 | 0.2271109903 | -6.7723700739E-007 | 0.2908362576 | 2.0858204730E-007 |
| 1.5 | 0.4823843992 | -4.7763943689E-007 | 0.8415768047 | 3.5292719458E-007 |
| 2.0 | 0.7949739993 | -2.5950480864E-007 | 1.6496506598 | 8.9354771737E-007 |
| 2.5 | 1.1356472829 | -1.1575548076E-007 | 2.6113910910 | 9.6941780958E-007 |
| 3.0 | 1.4811590655 | -4.6188424359E-006 | 3.6301217015 | 1.1506687414E-005 |
| 3.2 | 1.6165050544 | -0.00051449094 | 4.0396568830 | 0.0014489918 |

| ξ | | | | |
|---|---|---|---|---|
| 3.4 | 1.7076830315 | -0.04276749508 | 4.5893860584 | 0.1469808356 |
| 3.5 | 1.4627148414 | -0.35338134721 | 5.9789554323 | 1.3365109770 |

Table 2. Radius of convergence of the power series solution of TOV equation for the relativistic effect $\sigma = 0.2$ before doing acceleration.

| $\xi$ | $\theta(\xi_1)$-An | error ($\varepsilon_1$) | $\nu(\xi_1)$-An | error ($\varepsilon_2$) |
|---|---|---|---|---|
| 0.1 | 0.0031982938 | -2.2362586831E-006 | 0.0003326943 | -2.9907271354E-008 |
| 0.3 | 0.0286620788 | -2.2502026265E-006 | 0.0088465883 | 2.16251607697E-008 |
| 0.5 | 0.0789402709 | -2.1395188244E-006 | 0.0397404600 | 8.08164897295E-008 |
| 0.7 | 0.1527556329 | -1.9796703668E-006 | 0.1043317478 | 2.06352842008E-007 |
| 1.0 | 0.3034042901 | -1.6674736475E-006 | 0.2779523918 | 5.19178451952E-007 |
| 1.5 | 0.6393200909 | -1.0088775712E-006 | 0.7670949161 | 1.16339283773E-006 |
| 2.0 | 1.0404738378 | -3.5765433481E-007 | 1.4268048236 | 1.69882056111E-006 |
| 2.5 | 1.4623116228 | -2.6108234761E-006 | 2.1481527204 | 3.96874110464E-006 |
| 3.0 | 1.8307363050 | -0.0417696517 | 2.9183335680 | 0.05979638685 |
| 3.1 | 1.7175597721 | -0.2337994006 | 3.3725656715 | 0.37656844720 |
| 3.2 | 0.7921498711 | -1.2367731914 | 5.3648422145 | 2.23325307786 |

Table 3. Radius of convergence of the power series solution of TOV equation for the relativistic effect $\sigma = 0.3$ before doing acceleration.

| $\xi$ | $\theta(\xi_1)$-An | error ($\varepsilon_1$) | $\nu(\xi_1)$-An | error ($\varepsilon_2$) |
|---|---|---|---|---|
| 0.1 | 0.0041140941 | -2.3715336048E-006 | 0.0003325116 | -2.5482476363E-008 |
| 0.3 | 0.0368418883 | -2.4174640616E-006 | 0.0088033324 | 1.72283842661E-008 |
| 0.5 | 0.1013152851 | -2.3181737671E-006 | 0.0392122892 | 6.73057481138E-008 |
| 0.7 | 0.1955935387 | -2.1051340775E-006 | 0.1017001108 | 1.81159963003E-007 |
| 1.0 | 0.3864665335 | -1.6654161666E-006 | 0.2645710374 | 5.02286165494E-007 |
| 1.5 | 0.8034447178 | -8.0623353149E-007 | 0.6959310510 | 1.12674301400E-006 |
| 2.0 | 1.2834657367 | 7.36378646948E-009 | 1.2331174045 | 1.38017677043E-006 |
| 2.5 | 1.7667238793 | 0.00088563654 | 1.7804026764 | -0.0005676831 |
| 2.6 | 1.8657341938 | 0.00682973775 | 1.8819879809 | -0.0051201240 |
| 2.7 | 1.9989280051 | 0.04858109323 | 1.9493098672 | -0.0423219728 |
| 2.8 | 2.3605847415 | 0.32055375109 | 1.7713173023 | -0.3231297424 |

Therefore, it was necessary to improve the power series solution of TOV relativistic equation, which in turn contributes in obtaining a better approximation to the physical parameters of stars. A combination of the two techniques for Euler-Abel transformation and Padé approximation Nouh (2004) and Nouh & Saad (2013) has utilized. Figs. 1, 2 and 3 represent the solution curves of TOV equation in $\xi$-$\theta$ phase plane for the general relativistic effects $\sigma = 0.1$, $0.2$ and $0.3$ respectively. Each figure displays both analytical (dashed line) and numerical (solid line) solutions and the relative error. The maximum relative error was of order $10^{-3}$ for $\sigma = 0.1$, and $0.2$, while for $\sigma = 0.3$ the relative error was of order $10^{-2}$. By the same way, Figs. 4, 5 and 6 show the solution curves $\xi$-$v$ phase plane for the general relativistic effects $\sigma = 0.1$, $0.2$ and $0.3$ respectively. The most important finding is that the physical range of the convergent power series solutions extended to around ten times than of the classical procedures.

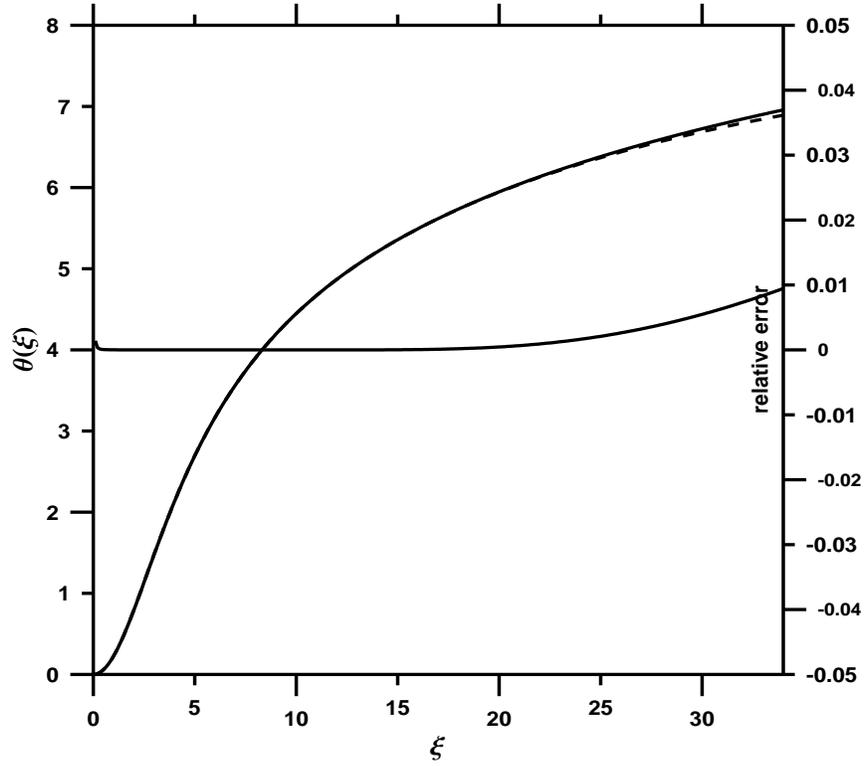

Fig. 1. Solution of TOV equation in $\xi$-$\theta$ phase plane for general relativistic effect $\sigma = 0.1$. Comparison between analytical (dashed) and numerical (solid) solutions gives a maximum relative error of order $10^{-3}$.

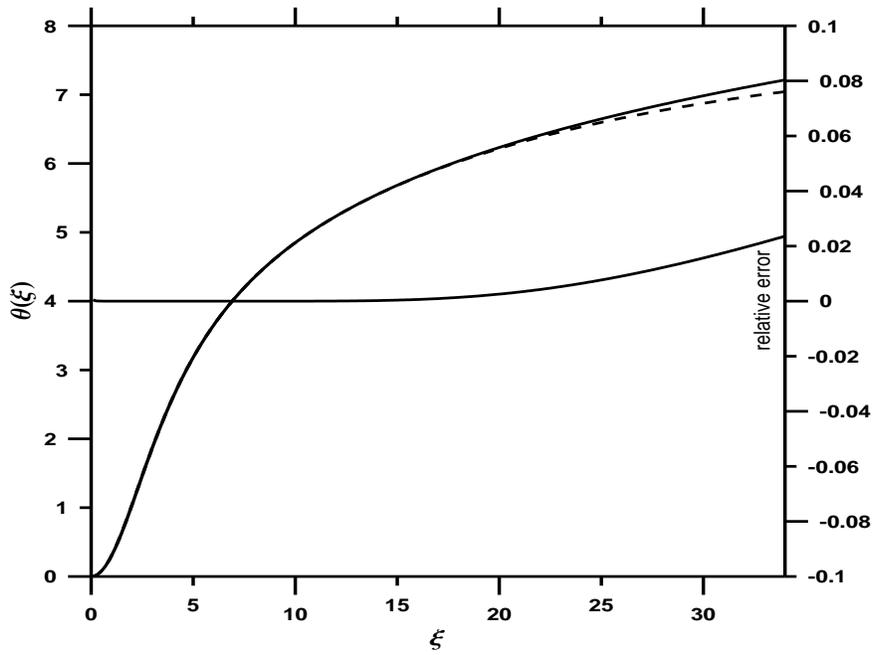

Fig. 2. Solution of TOV equation in $\xi$-$\theta$ phase plane for general relativistic effect $\sigma = 0.2$. Comparison between analytical (dashed) and numerical (solid) solutions gives a maximum relative error of order $10^{-3}$.

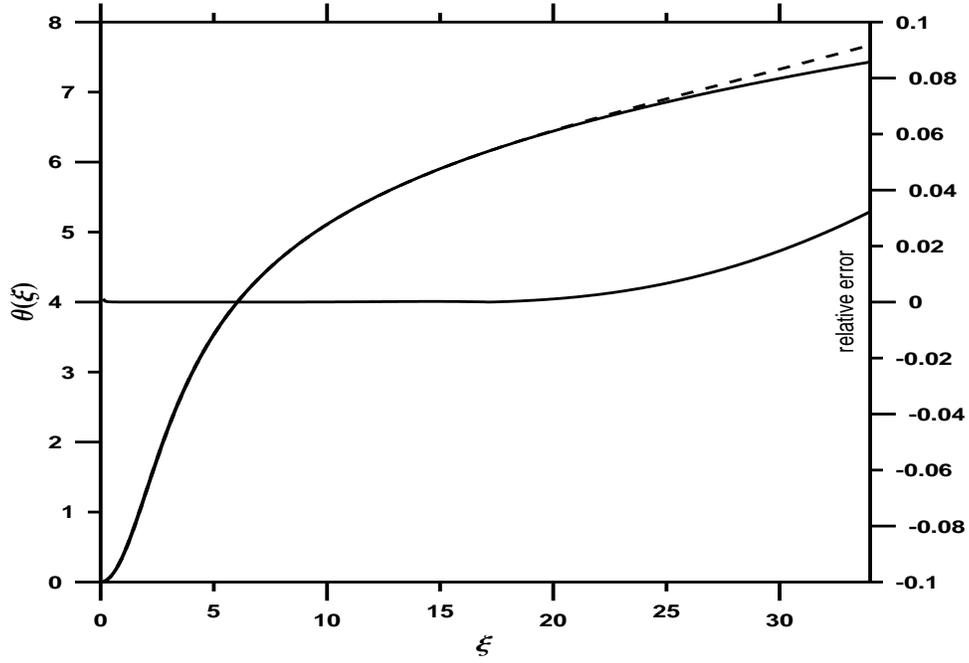

Fig. 3. Solution of TOV equation in $\xi$-$\theta$ phase plane for general relativistic effect $\sigma = 0.3$. Comparison between analytical (dashed) and numerical (solid) solutions gives a maximum relative error of order $10^{-2}$.

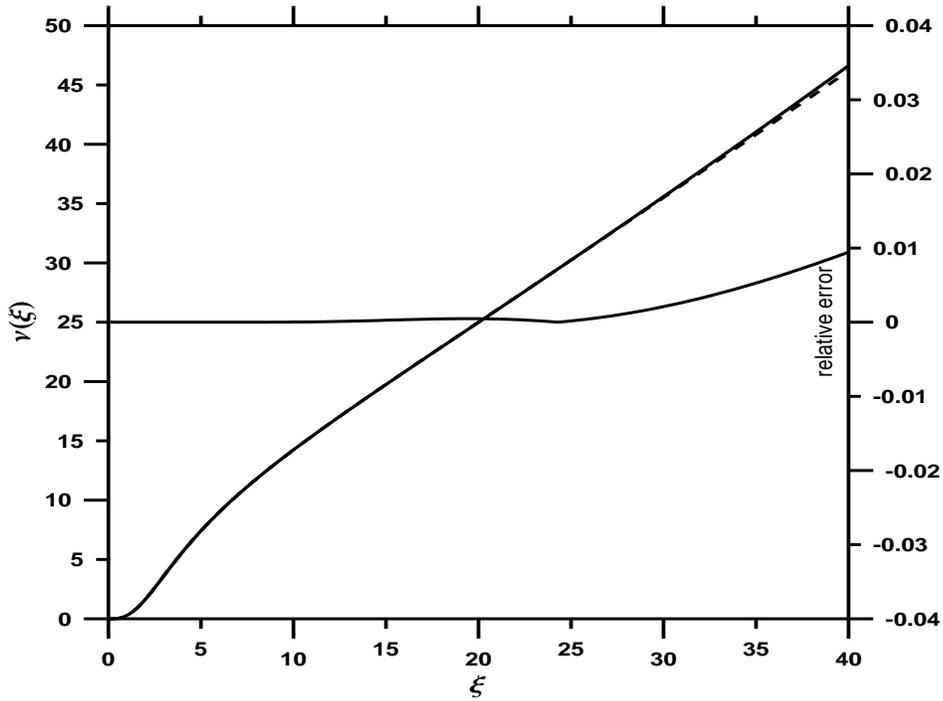

Fig. 4. Mass function $v(\xi)$ in $\xi$-$v$ phase plane for general relativistic effect $\sigma = 0.1$. Comparison between analytical (dashed) and numerical (solid) solutions gives a maximum relative error of order $10^{-3}$.

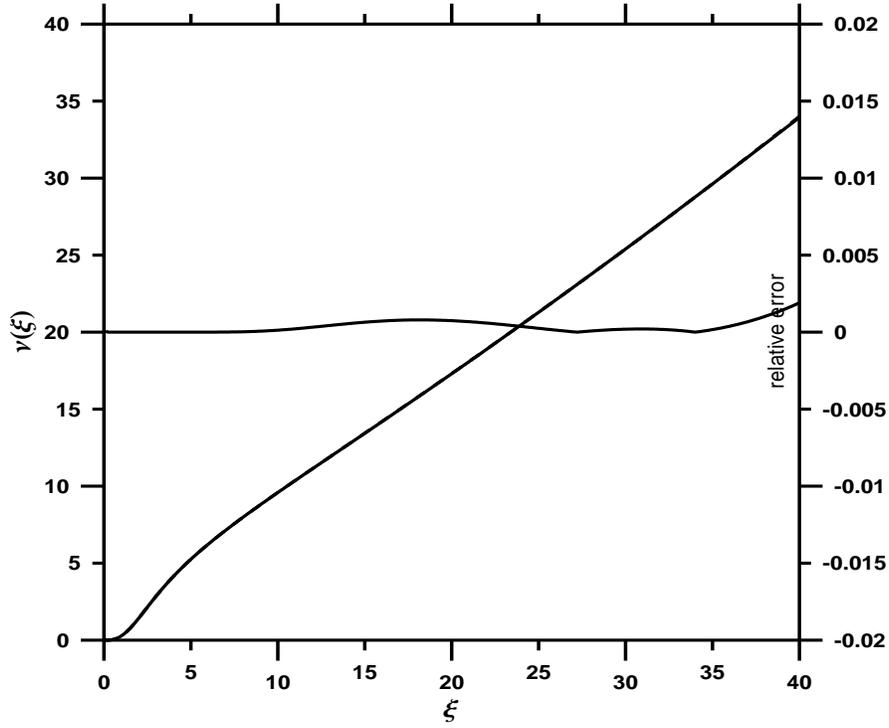

Fig. 5. Mass function $v(\xi)$ in $\xi$-$v$ phase plane for general relativistic effect $\sigma = 0.2$. Comparison between analytical (dashed) and numerical (solid) solutions gives a maximum relative error of order $10^{-3}$.

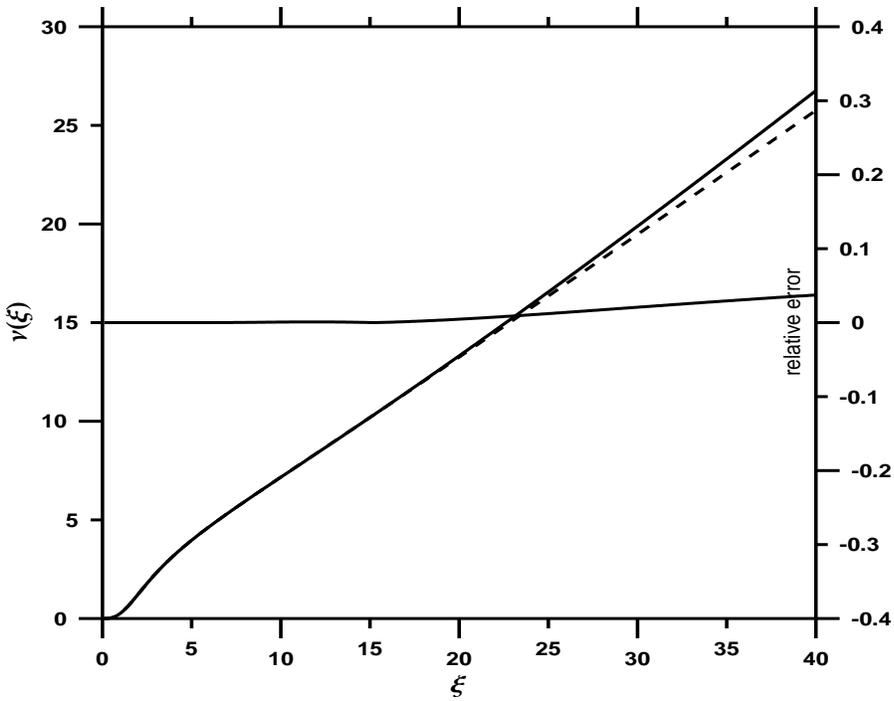

Fig. 6. Mass function $v(\xi)$ in $\xi$-$v$ phase plane for general relativistic effect $\sigma = 0.3$. Comparison between analytical (dashed) and numerical (solid) solutions gives a maximum relative error of order $10^{-2}$.

## 4.1. Application to Neutron Stars

The aim of the study of stellar structure, either in the framework of Newtonian mechanics (non-relativistic) or in the framework of general relativity (relativistic) is to investigate and examine the behavior of the physical parameters, such as pressure, density, temperature, mass effective, etc., and determine the mass-radius relation. In this context, the proposed analytical solution in the present paper has applied to a neutron star of physical parameters: mass $M = 1.5 M_\odot$, central density $\rho_c = 5.75 \times 10^{14} \, g \, cm^{-3}$, pressure $P = 2 \times 10^{33}$ par, and radius $R = 1.4 \times 10^6 \, cm$. Tables 4, 5 and 6 show the physical parameters of a neutron star coming from a direct analytical solution of TOV equation with the relativistic isothermal configurations $\sigma = 0.1, 0.2$ and $0.3$. The tables give the mass ratio $M/M_0$ in the first column and its error between analytical and numerical $\varepsilon_1$, while columns 3 and 4 show the density ratio $\rho/\rho_c$ and its error $\varepsilon_2$. The error is increasing gradually and the radius of convergence to the power series solution is limited. Therefore, the calculated physical parameters have poor accuracy

On the other hand, manipulation of the analytical solution with the utility of the proposed procedure improved the solution, which of course will be reflected on the accuracy of the estimated physical parameters. Figs. 7, 8 and 9 describe the density profile for isothermal spheres with relativistic effects $\sigma = 0.1, 0.2$ and $0.3$. As shown, the density decreases with increasing the radius of the star. In all the figures the analytical and numerical curves cannot be distinguished. Comparison between analytical (solid) and numerical (dashed) solutions gives a maximum relative error of order $10^{-4}$. Figs. 10, 11 and 12 give an analytical and numerical description to the ratios $M/M_0$ and their errors.

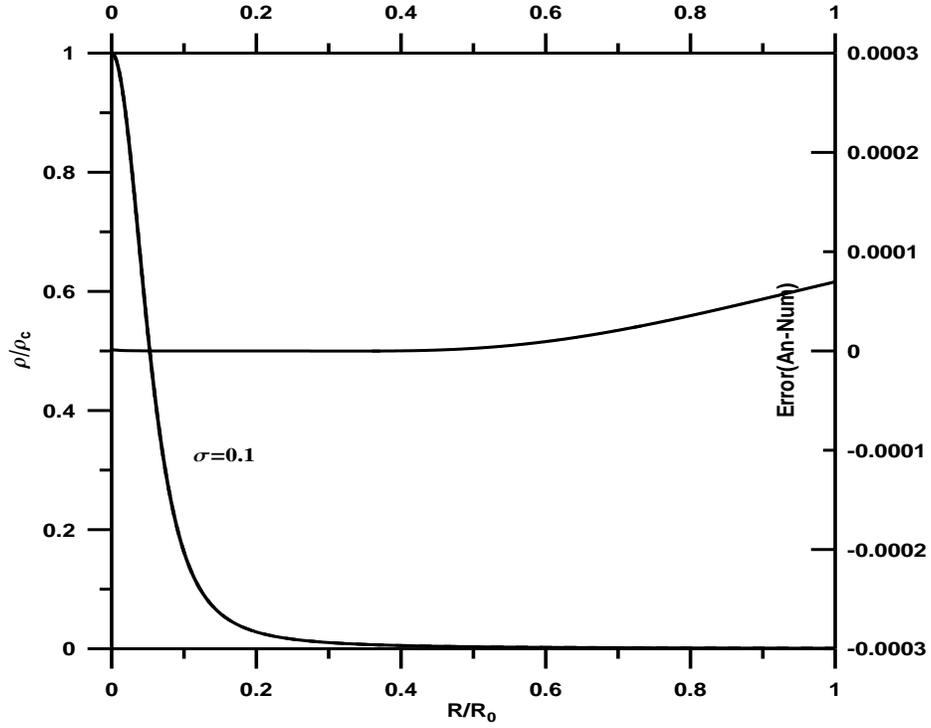

Fig. 7. Density profile for isothermal spheres with relativistic effects $\sigma=0.1$. Comparison between analytical (solid) and numerical (dashed) solutions gives a maximum relative error of order $10^{-4}$.

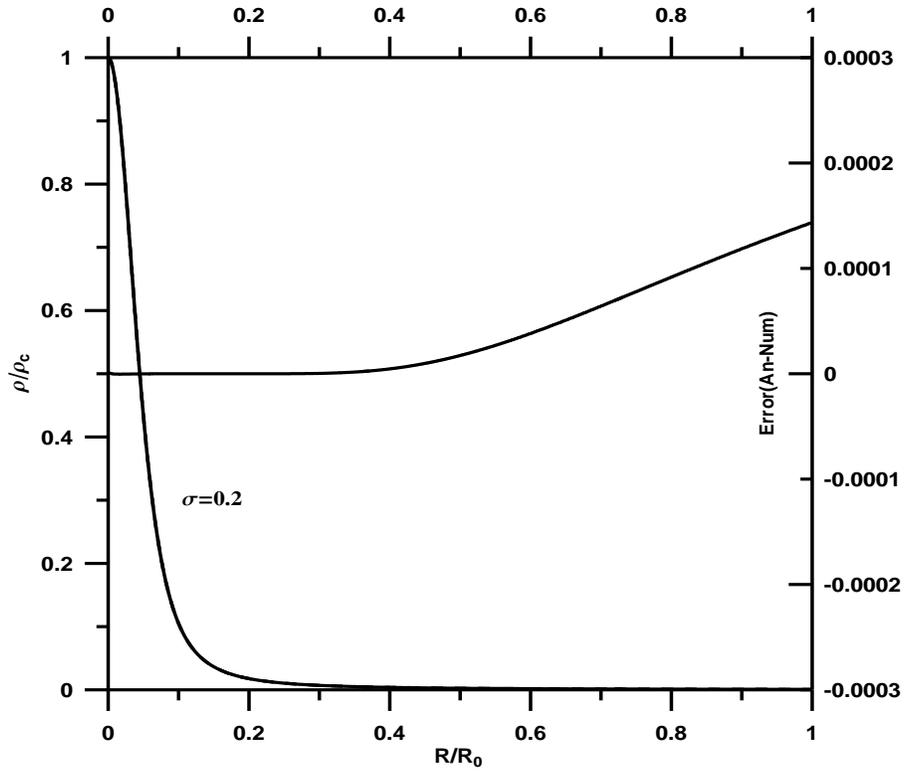

Fig. 8. Density profile for isothermal spheres with relativistic effects $\sigma=0.2$. Comparison between analytical (solid) and numerical (dashed) solutions gives a maximum relative error of order $10^{-4}$.

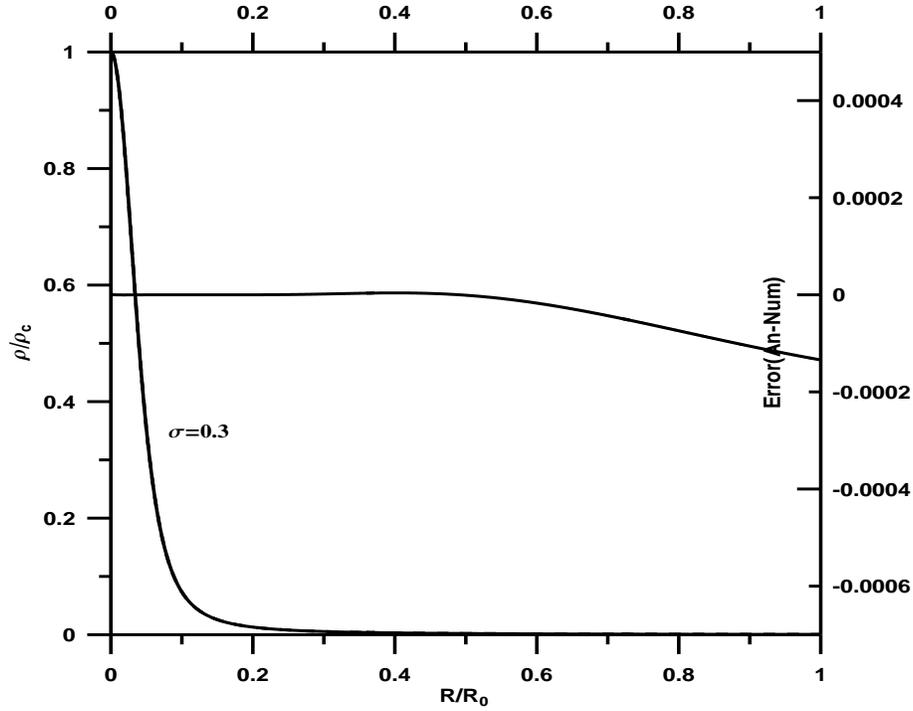

Fig. 9. Density profile for isothermal spheres with relativistic effects $\sigma=0.3$. Comparison between analytical (solid) and numerical (dashed) solutions gives a maximum relative error of order $10^{-4}$.

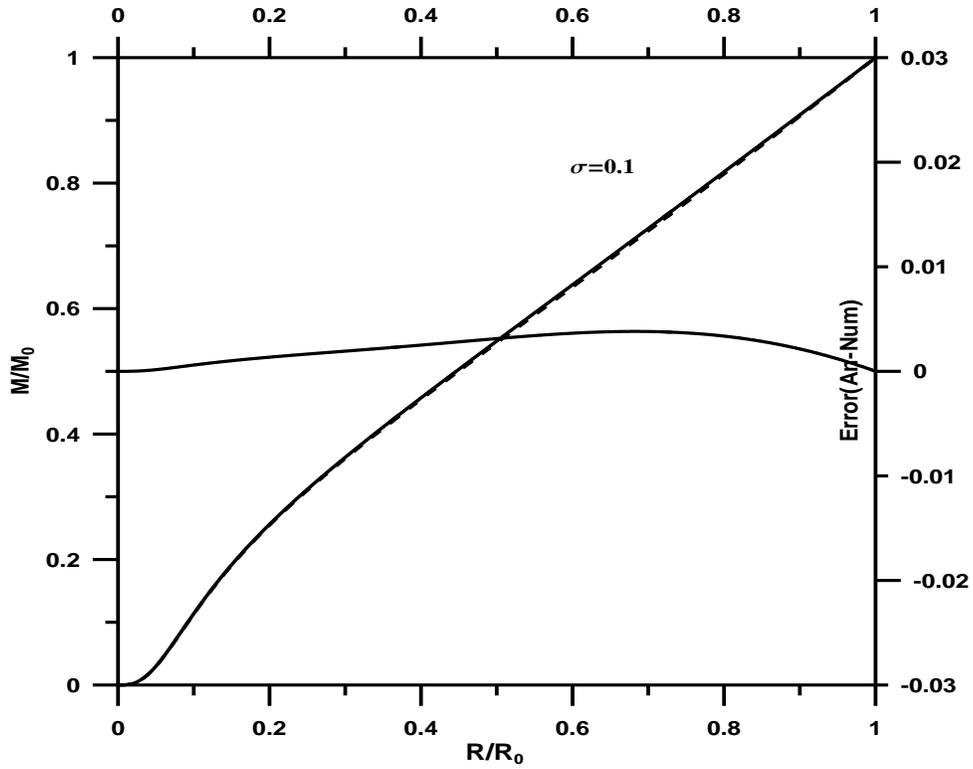

Fig. 10. Mass profile for isothermal spheres with relativistic effects $\sigma=0.1$. Comparison between analytical (solid) and numerical (dashed) solutions gives a maximum relative error of order $10^{-3}$

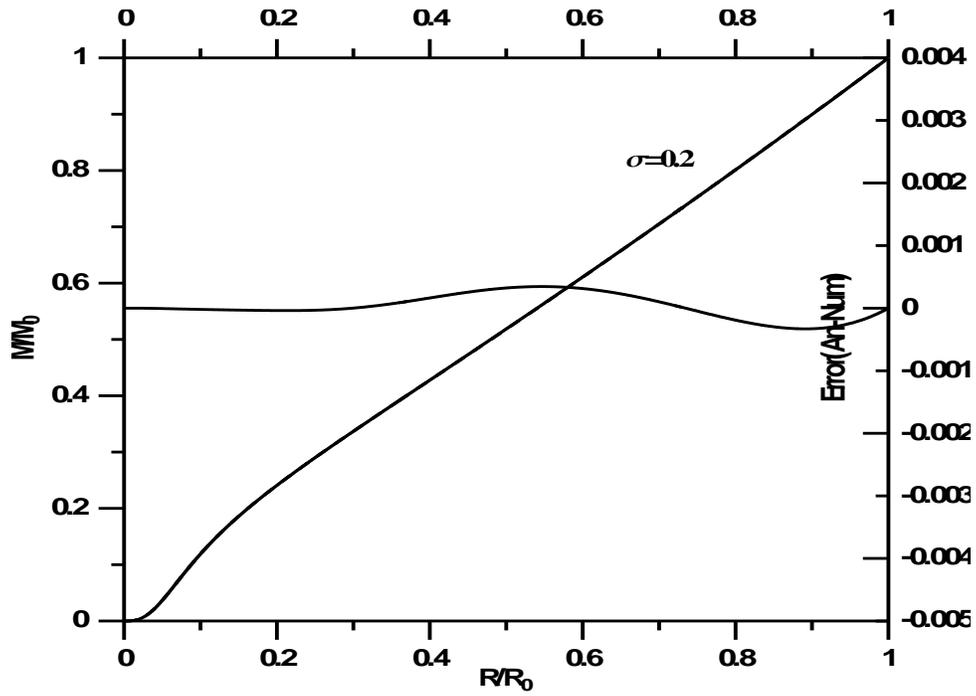

Fig. 11. Mass profile for isothermal spheres with relativistic effects $\sigma=0.2$. Comparison between analytical (solid) and numerical (dashed) solutions gives a maximum relative error of order $10^{-3}$

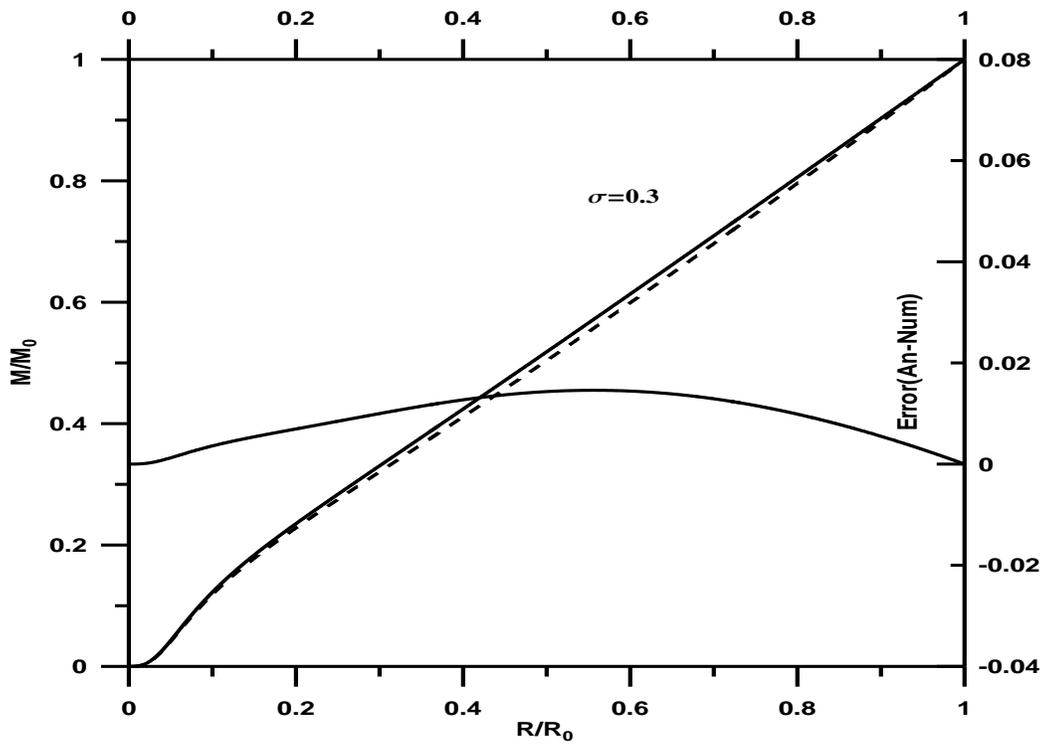

Fig. 12. Mass profile for isothermal spheres with relativistic effects $\sigma=0.2$. Comparison between analytical (solid) and numerical (dashed) solutions gives a maximum relative error of order $10^{-2}$.

Table 4. Physical parameters of a neutron star before accelerating the power series solution with the relativistic isothermal configuration $\sigma = 0.1$

| $\xi$ | $M/M_0$-An | error ($\varepsilon_1$) | $\rho/\rho_c$-An | error ($\varepsilon_2$) |
|---|---|---|---|---|
| 0.1 | 6.50157995E-005 | -7.93658100E-006 | 0.99762067 | 1.02359751E-006 |
| 0.3 | 0.00173556 | -0.00021188 | 0.97887096 | 8.54561849E-007 |
| 0.5 | 0.00785616 | -0.00095910 | 0.94283984 | 7.77213760E-007 |
| 0.7 | 0.02085504 | -0.00254606 | 0.89223216 | 6.87572904E-007 |
| 1.0 | 0.05680799 | -0.00693531 | 0.79683234 | 5.39644164E-007 |
| 1.5 | 0.16438215 | -0.02006837 | 0.61730972 | 2.94851398E-007 |
| 2.0 | 0.32222029 | -0.03933779 | 0.45159298 | 1.17190534E-007 |
| 2.5 | 0.51007357 | -0.06227168 | 0.32121414 | 3.71822947E-008 |
| 3.0 | 0.70905853 | -0.08656227 | 0.22737399 | 1.05020223E-006 |
| 3.2 | 0.78905155 | -0.09601309 | 0.19859155 | 0.00010215 |
| 3.4 | 0.89642816 | -0.07722548 | 0.18128534 | 0.00758967 |
| 3.5 | 1.16784772 | 0.150350954 | 0.23160665 | 0.06894714 |

Table 5. Physical parameters of a neutron star before accelerating the power series solution with the relativistic isothermal configuration $\sigma = 0.2$

| $\xi$ | $M/M_0$-An | error ($\varepsilon_1$) | $\rho/\rho_c$-An | error ($\varepsilon_2$) |
|---|---|---|---|---|
| 0.1 | 1.84156019E-010 | -5.01634030E-012 | 0.99999968 | 3.01749311E-008 |
| 0.3 | 0.00148393 | -1.90493760E-009 | 0.98718236 | 2.24118632E-006 |
| 0.5 | 0.01151679 | 6.649477030E-009 | 0.95026265 | 2.08770747E-006 |
| 0.7 | 0.03735025 | 3.813366056E-008 | 0.89281237 | 1.84370015E-006 |
| 1.0 | 0.11605026 | 2.927041631E-008 | 0.77975328 | 1.39542662E-006 |
| 1.5 | 0.36081190 | 9.344106428E-008 | 0.56773599 | 6.57116908E-007 |

| $\xi$ | | | | |
|---|---|---|---|---|
| 2.0 | 0.71167074 | 1.328278937E-007 | 0.38361304 | 1.87786358E-007 |
| 2.5 | 1.10745911 | 7.846480399E-008 | 0.25185372 | 4.60871126E-008 |
| 3.0 | 1.50812170 | 0.005015086 | 0.16761343 | 0.00119469 |
| 3.1 | 1.61367473 | 0.033657910 | 0.16028642 | 0.00667063 |
| 3.2 | 1.86778757 | 0.211838462 | 0.18008080 | 0.03811105 |

Table 6. Physical parameters of a neutron star before accelerating the power series solution with the relativistic isothermal configuration $\sigma = 0.3$

| $\xi$ | $M/M_0$-**An** | error ($\varepsilon_1$) | $\rho/\rho_c$-**An** | error ($\varepsilon_2$) |
|---|---|---|---|---|
| 0.1 | 3.38314717E-010 | -6.88539237E-012 | 0.99999959 | 2.92341651E-008 |
| 0.3 | 0.00272011 | -9.11526077E-009 | 0.98354708 | 2.44318533E-006 |
| 0.5 | 0.02097454 | 1.422439966E-008 | 0.93656827 | 2.23634834E-006 |
| 0.7 | 0.06731848 | 4.124582367E-008 | 0.86471217 | 1.91681736E-006 |
| 1.0 | 0.20465838 | 1.439727668E-007 | 0.72796240 | 1.32236526E-006 |
| 1.5 | 0.60732845 | 2.002381181E-007 | 0.48984974 | 4.83381012E-007 |
| 2.0 | 1.14039429 | -1.15562337E-007 | 0.30525537 | 4.27123615E-008 |
| 2.5 | 1.69940711 | -6.07445250E-005 | 0.18780008 | -2.0258932E-005 |
| 2.6 | 1.80808156 | -0.000591672 | 0.17072850 | -0.0001545 |
| 2.7 | 1.91106920 | -0.005312332 | 0.15461843 | -0.0010811 |
| 2.8 | 1.97859450 | -0.043856014 | 0.13523016 | -0.0068658 |
| 2.9 | 1.79221566 | -0.334569026 | 0.09371845 | -0.0361909 |

## 5. Conclusions

We have constructed general analytical formulations for solving the hydrostatic equilibrium equation (TOV equation) in the framework of relativistic isothermal gas spheres. Traditional procedures solutions were not effective and the radius of convergence of the power series solution was limited as shown in Tables 1, 2 and 3. That is the maximum radii of convergence for the relativistic effects $\sigma = 0.1, 0.2$ and $0.3$ are $\xi_1 = 3.58$, $\xi_1 = 3.23$ and $\xi_1 = 2.8$ respectively. When we improved the power series solution with utilizing a combination of the two techniques for Euler-Abel transformation and Padé approximation (Nouh 2004) and Nouh & Saad (2013), the radius of convergence increased. As a result, the physical range of the convergent power series solutions extended to about ten times than of the traditional procedures ($\xi_1 \approx 35$) as displayed in Figs. 1, 2 and 3. The implementation of the proposed analytical solution contributed in improving the accuracy of the calculation of the physical parameters of the stars. As it is shown in Figs. 7 to 12 for density and mass profiles of the star, the maximum relative error reached $10^{-3}$, which establishes the validity and accuracy of the suggested procedure. The derived analytical solution in the present paper may be treated using more convenience methods, which in turn will be a good contribution in the field of the stellar structure.